\documentclass{article}

\usepackage{arxiv}

\usepackage[utf8]{inputenc} 
\usepackage[T1]{fontenc}    
\usepackage{hyperref}       
\usepackage{url}            
\usepackage{booktabs}       
\usepackage{amsfonts}       
\usepackage{nicefrac}       
\usepackage{microtype}      
\usepackage[usenames,dvipsnames]{color}
\usepackage{caption}
\usepackage{cleveref}
\usepackage{graphicx}
\usepackage{lmodern,babel,adjustbox,booktabs,multirow}
\usepackage[numbers]{natbib}
\usepackage{subcaption}
\usepackage{booktabs}
\usepackage{multirow}
\usepackage{amssymb}
\usepackage{pifont}
\newcommand{\cmark}{\ding{51}}%
\newcommand{\xmark}{\ding{55}}%
\usepackage{authblk}

\definecolor{LightGreen}{cmyk}{1.0, 1.0, 1.0, 1.0}

\title{TextileNet: A Material Taxonomy-based Fashion Textile Dataset}

\author[1]{Shu Zhong}
\author[2]{Miriam Ribul}
\author[1]{Youngjun Cho}
\author[1]{Marianna Obrist}
\affil[1]{Department of Computer Science, University College London}
\affil[2]{Materials Science Research Centre, Royal College of Art}

\begin{document}

\maketitle

\begin{abstract}
  The rise of Machine Learning (ML) is gradually digitalizing and reshaping the fashion industry. Recent years have witnessed a number of fashion AI applications, for example, virtual try-ons. Textile material identification and categorization play a crucial role in the fashion textile sector, including fashion design, retails, and recycling. 
  At the same time, Net Zero is a global goal and the fashion industry is undergoing a significant change so that textile materials can be reused, repaired and recycled in a sustainable manner. 
  There is still a challenge in identifying textile materials automatically for garments, as we lack a low-cost and effective technique for identifying them. In light of this, we build the first fashion textile dataset, TextileNet, based on textile material taxonomies - a fibre taxonomy and a fabric taxonomy generated in collaboration with material scientists. TextileNet can be used to train and evaluate the state-of-the-art Deep Learning models for textile materials. We hope to standardize textile related datasets through the use of taxonomies. TextileNet contains 33 fibres labels and 27 fabrics labels, and has in total of 760,949 images. We use standard Convolutional Neural Networks (CNNs) and Vision Transformers (ViTs) to establish baselines for this dataset.
  Future applications for this dataset range from textile classification to optimization of the textile supply chain and interactive design for consumers. We envision that this can contribute to the development of a new AI-based fashion platform.
\end{abstract}

\section{Introduction}

Clothing and textiles are ubiquitous in our daily lives. Online shopping elevates individuals to a new level of purchasing experience - customised shopping with tons of options from a global market, simple checkout procedures, convenient delivery and returns. According to market estimates, the global fashion e-market is worth \$752.5 billion in 2020  \cite{coppola2021commerce}. 
This enormous economic value indicates an increased demand for e-commerce services by individuals. These rising demands in the fashion industry motivate the use of Machine Learning (ML) techniques to facilitate low-level pixel recognition, mid-level fashion comprehension and high-level fashion applications \cite{song2018multimedia}. 

Higher-level applications, such as outfit recommendations and virtual try-ons \cite{han2018viton}, are supported by lower-level fashion tasks, \textit{e.g.} parsing (segmentation) \cite{ge2019deepfashion2,jia2020fashionpedia}, landmark detection \cite{liu2016deepfashion}, \textit{etc.}  A number of works have then developed apparel related datasets for all levels of fashion tasks, including landmark annotations \cite{liu2016deepfashion,jia2020fashionpedia,ge2019deepfashion2}, category classification \cite{guo2019imaterialist,zou2019fashionai,xiao2017fashion}, attributes labelling \cite{jia2020fashionpedia,he2016deep}, recognition-based retrieval \cite{huang2015cross,han2018viton,hadi2015buy}, \textit{etc.}

Despite the development of ML techniques in the fashion industry, the textile industry still faces challenges in its chase of a more sustainable model to reduce the enormous volumes of textile wastes and to meet the global Net Zero goal \cite{macarthurfashion,boiten2017circular}. 
Textile materials play a critical role in garments due to the fact that it is selected based on their particular properties, which may include the level of comfort they provide and the degree to which they can be recycled, \textit{etc.} \cite{laitala2012sustainable}. Millions of tons of garments end up in landfill every year \cite{muthu2018circular}. Textiles circularity, a novel conceptual model of the circular economy, demonstrates an option for the fashion industry to reduce its carbon footprint and costs, maximise the life of textiles, and minimise waste. Textiles can be reused at many levels, being regenerated into new fibres or utilised textile wastes as energy fuel \textit{etc.}, thereby reducing the carbon footprint \cite{macarthur2013towards}. It is a recommended practice to ensure that textiles can be traced back to their original source so that the recycling process can be guaranteed. 

Yet, nowadays, textiles are mostly sorted manually \cite{norup2019evaluation}, despite recent research raise using near infrared spectroscopy (NIR spectroscopy) to recognize textiles for automated garments sorting line \cite{cura2021textile}. \textbf{A low-cost, high-efficiency technique for the automatic identification of textile materials in garments is missing}, so that the digitised fashion sector would be able to retrieve the materials they are composed of. This would help reduce a large amount of textile wastes and carbon emissions \cite{textileexchange2021preferred}. 

Given the significance of textile material identification in clothing, it is worth noticing that this identification process can be complicated because fibre and fabric refer to different textile materials. Fibre is the material to make fabric, however, most existing fashion datasets in ML mixed them in the same class. No dataset presently contains organized textile material labels; they do not provide a systematic picture of material-related labels \cite{liu2016deepfashion,guo2019imaterialist}. They contain partial textile material attributes; their annotation scheme lacks a rationale and is not systematically reviewed by material scientists. 
\begin{figure}
  \centering
    \centering
    \includegraphics[width=0.9\textwidth]{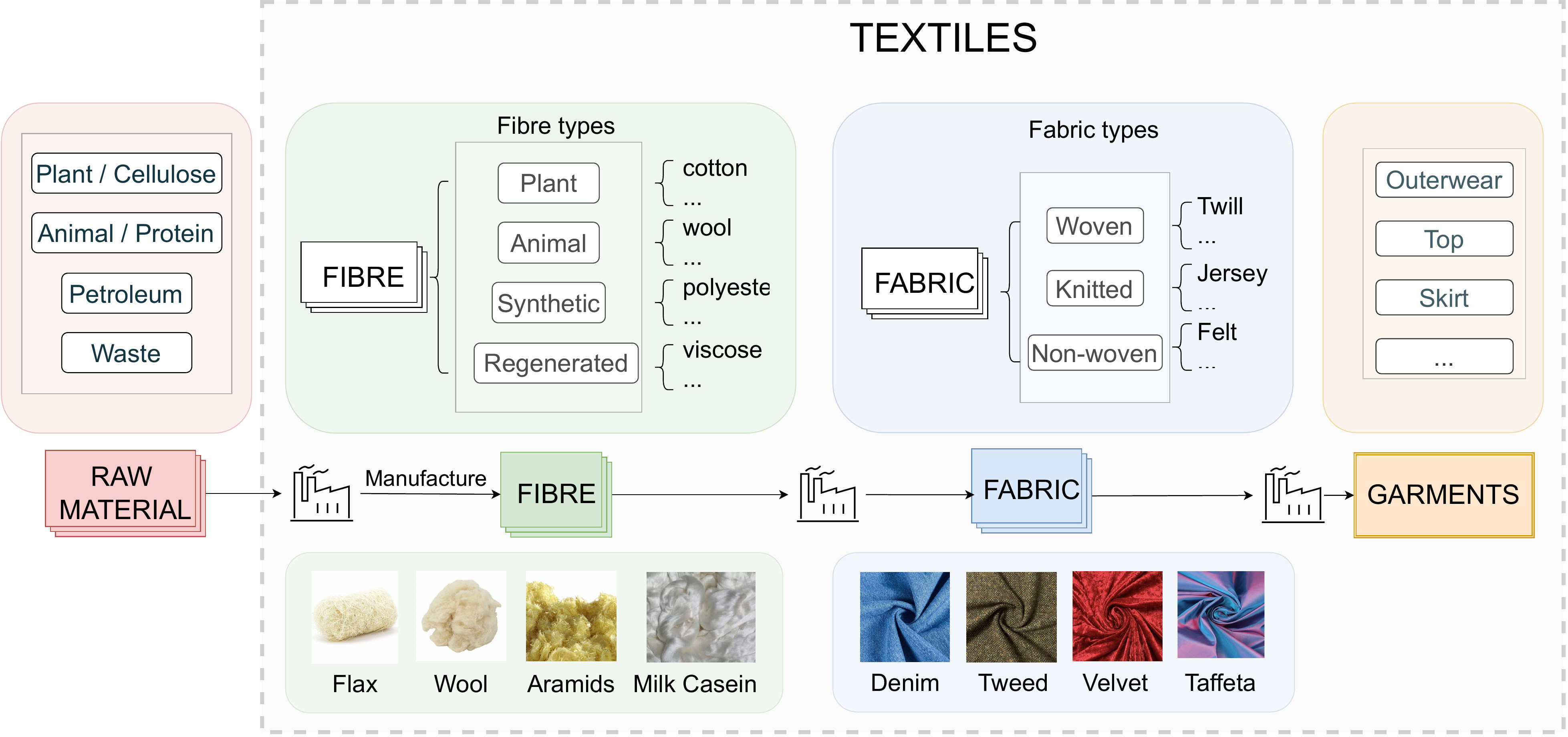}
    \captionof{figure}{The general production flow of textiles. \textit{Textiles} is an umbrella term and it includes raw material, fibre, fabric and garments \textit{etc} (indicated in dashed box). The dataset labels are generated from two textile material taxonomies: a fibre taxonomy (\Cref{fig:fibre}) and a fabric taxonomy (in Supplementary material). Fibres have four macro-types and we show several fibre examples (flax, wool, aramids, and milk casein) in the figure. Fabrics have three macro-types of production methods, we also show several fabric examples (denim, tweed, velvet and taffeta) .}
    \label{fig:textile}
    \vspace{-15pt}
\end{figure}

Here we propose TextileNet, a material taxonomy-based fashion textile dataset to close the gap in current research on textile material identification in clothing. We developed the TextileNet dataset (illustrated in \Cref{fig:textile} and the detailed illustration is in \Cref{fig:fibre}) based on fibre and fabric labels. We achieved this through carefully designing two taxonomies, one for fibres and another for fabrics in close collaboration with material domain experts. We discuss the design details in \Cref{sec:fashion}.

The goal of this TextileNet dataset is to contribute to textile material identification in the fashion industry and image-based textile material retrieval, at the same time, \textbf{standardize the digitized textile material labelling}. 
TextileNet can be deployed in various domains, including material science, fashion design, retails and the textile supply chain, \textit{etc}. Our contributions are:

\begin{itemize}
    \item We present a fibre taxonomy and a fabric taxonomy created in collaboration with material domain experts; these taxonomies contain macro-types of textiles and are extendable for future new fibre/fabric types;
    \item Using the labels from these taxonomies, we collect and build material taxonomy-based fashion datasets for fibre and fabric. The built datasets, named TextileNet, TextileNet-fibre contains 33 fibre labels, 27 fabric labels in TextileNet-fabric, and have 760,949 images;
    \item We present and report two baseline models (CNNs and Vision Transformers) for fibre and fabric classification, both models achieve $>80\%$ top-5 accuracy on our datasets.
\end{itemize}

\section{Related work}
\subsection{Why do we care about fibre, fabric and textiles?}
\label{sec:related:why}
The general production flow in the fashion industry contains the following intermediate stages: \textit{raw material}, \textit{fibre}, \textit{fabric} and \textit{garment} (as illustrated in \Cref{fig:textile}). The term textiles is commonly used as an umbrella term to describe any textile-related products; while fibre normally refers to the raw elements or materials that is used to manufacture fabric. In other words, fabric is a finished product made from fibre, and the term is defined by its manufacturing processes rather than to its constituent material (fibre). Velvet, for example, is a type of woven tufted fabric with evenly distributed cut threads and a short pile that can be fabricated from silk, rayon, and polyester \textit{etc}. We present a detailed definition of fibre, fabric and textile in \Cref{sec:textile} to clarify the difference between these concepts. Here we first elaborate on the importance of this research in the broader context of a circular economy.

Recent years have witnessed the fashion industry, including brands and retailers, embrace the need for creating more sustainable fashion. Brands such as Stella McCartney, Nike, Doodlage \textit{etc.} have used recycled materials in their products and created recycling programmes for their customers \cite{khandual2019fashion}. Researchers developed bio-degradable techniques for existing textiles and created novel sustainable fibres \cite{ribul2021mechanical}. Multiple scientific domains, from material science, supply chain, to consumer experience, are contributing to the circular economy of sustainable textiles futures.

Improving the sustainability of the fashion industry requires a comprehensive understanding of the fibres and fabrics used in garments. If we can track the origin of the raw material used in clothes, we can generate more accurate carbon footprint models for them \cite{macarthur2013towards} and material experts can begin to research and design novel recycling processes for garments \cite{laitala2012sustainable}. In a circular model, all textile materials are considered as resources, hence there is no waste \cite{textilecircularity}. Our TextileNet dataset can be seen as the first step in automating the process of tracing back the fibre and fabric origins of garments, linking together all steps in the textiles production flow. 

\subsection{Textile classification}

Traditional material classifications in material science typically involve physical, chemical, or biological tests. \textcolor{LightGreen}{Classifications tasks are often assigned for fibres, whereas fabric-related tasks typically focus on defect detections.}

\paragraph{Fibre} The infrared spectroscopy data were used to identify and classify textile fibres. Peets \textit{et al.} \cite{peets2017identification} used Principal Component Analysis (PCA) to analyse spectral data in chemometric methods, and only silk and wool could be distinguished partially. For textile fibre recognition tasks, all research is limited by sample size (a few hundred samples). Liu \textit{et al.} \cite{liu2020qualitative} demonstrated that CNN outperforms Support Vector Machines (SVM) and Multilayer Perceptron (MLP) algorithms with 263 samples tested from the near infrared spectroscopy (NIR) spectrum ; whereas Riba \textit{et al.} \cite{riba2020circular} integrated Attenuated total reflectance-Fourier transform infrared (ATR-FTIR) spectra with ML models to test 350 samples in 7 types of fibres (all included in our fibre taxonomy), achieving 100\% accuracy. 
In recognising cashmere and wool fibres, microscopic pictures coupled with ML approaches have proven to be effective \cite{xing2022application, zhong2017wool}. \textcolor{LightGreen}{We further compare these methods in our supplementary materials.}

\paragraph{Fabric} Morphological analysis is widely used in fabric classification and defect detection \cite{zhang1995fabric,huang2000woven}. Longhini \textit{et al.} \cite{longhini2021textile} investigated how robotic actions, such as pulling and twisting of textiles, can be used to classify fabrics that are made of wool, cotton or polyester fibre but with different production methods (woven or knitted). They also summarized a production method based taxonomy with 7 fibres (used only 3 in their analysis) and 11 fabrics (used 2 meta-types), this is the similar taxonomy that appeared in \textit{``Textiles and Fashion  Materials, Design and Technology''} by Sinclair \textit{et al.} \cite{sinclair2014textiles}. This taxonomy is small and incomplete, whereas our proposed taxonomy contains a greater number of fibre and fabric types and is based on their origins plus their production methods. Sun \textit{et al.} \cite{sun2016classification} used spectroscopy-based pattern recognition on textile fabrics. Fabric defect detection and classification using image analysis are discussed in \Cref{sec:fabricdataset}, these sample images are focused solely on pieces of fabric.

Our proposed TextileNet dataset looks at this textile retrieval and classification problem from a very different angle, the data we source is simple ubiquitous RGB images. Fetching such data in the wild would require simply a camera-based system, without the need of using any specialised equipment (\textit{eg.} spectrum analysis).

\subsection{Textile datasets}
\label{dataset}
Here we review existing image-based textile datasets both from the fashion and material domains. Fashion datasets include fashion items (e.g. garments), whereas material domain datasets focus on pieces of textile materials (e.g. fabric samples).

\paragraph{Fashion datasets}
\label{sec:fashion}

Fashion-related datasets were built for different purposes,  including fashion recommendation \cite{song2019gp,ni2019justifying}, categorization\cite{liu2016deepfashion,ge2019deepfashion2,zou2019fashionai,guo2019imaterialist},  3D Modeling \cite{bertiche2020cloth3d,madadi2021cloth3d++,korosteleva2021generating}, \textit{etc.} As large scale datasets are typically a result of a huge crawl of online fashion retail stores, the majority of these works employ the highest frequency metadata for image labelling \cite{liu2016deepfashion}. In other words, these datasets create and use labels based on what was presented on these online fashion websites. Since these dataset labels are derived from the fashion website where the images are collected; labels can be influenced by the source and fashion trends \cite{holland2018dataset}.

Despite the fact that these fashion datasets are multi-task datasets, a significant portion of their images lack material annotations. DeepFashion is the most popular fashion dataset and has the most extensive scale labels (1000 labels) generated by crawled online information \cite{liu2016deepfashion}. There are 218 labels in the class `fabric' among the thousand labels. Nonetheless, a great number of these labels are either duplicated or are fibres or may not even be textile materials. Take the first 10 listed fabric labels as an example: ``acid, acid wash, applique, bead, beaded, beaded chiffon, beaded sheer, bejeweled, bleach, bleached". Whereas acid is a fibre, bead is a round piece of plastic, acid wash and bleach are neither fibre nor fabric: they are textile manufacturing processes. In addition, certain essential fashion components, such as wool which had the largest market share in animal fibres \cite{textileexchange2021preferred}, are absent. 

Some fashion datasets contain textile material related labels, however, it is unclear how these labels are constructed, textile materials are labelled in the class either \textit{fabric} \cite{liu2016deepfashion, ge2019deepfashion2} or \textit{material} \cite{guo2019imaterialist, zou2019fashionai}. 
Fabric and fiber are in the same class named `Fabrics', and these labels do not aid in identifying the precise raw material of the clothing, making it difficult to distinguish between fibres and fabrics and track their origin for recycling. For instance, if we consider the iMaterialist Fashion Attribute Dataset \cite{guo2019imaterialist}, the first million level fashion dataset, $58.4\%$ of the dataset contains material/textile information, however, some images were labelled with multiple `material' labels without a clear definition of whether they are fibres or fabrics or even raw materials. \textcolor{LightGreen}{Another case is FashionAI \cite{zou2019fashionai}, they mentioned in the paper that they have material labels, such as \textit{cotton} and \textit{denim} in the material category, whereas denim is a fabric made from the fibre cotton. Unfortunately, this dataset is not fully disclosed to the public and we are unable to have further discussion.}


\paragraph{Material datasets}
\label{sec:fabricdataset}
Apart from fashion-related datasets, there are several domain-specific datasets created by material scientists. \textcolor{LightGreen}{It is worth mentioning that the objective for this dataset is fashion textiles, works related to surface texture indoors and in the wild, such as Materials in Context Database (MINC) \cite{bell2015material}, were excluded from further discussion.} Existing fashion textile material datasets are small and focus solely on the fibre or fabric level without a full picture of the fashion item: Kampouris \textit{et al.} developed the Fabrics Dataset that consists of micro-geometry fine-grained fabric surfaces \cite{kampouris2016fine}; Chetverikov \textit{et al.} \cite{chetverikov2002finding} and Bissi \textit{et al.} \cite{bissi2013automated} used the Textile Texture Database (TILDA) \cite{groupworkgroup} for textile quality inspection. FabricID \cite{wang2018fabric} and Fabric 1.0 \cite{shen2019large} employed fabric images with Deep Learning methods to identify fabrics, however, they both used self-coded names in the dataset rather than the actual names of fabrics.

\begin{table}[ht]
\centering
\caption{Comparison between TextileNet and the existing datasets with material labels. Material Labels* show the number of labels that are in both the dataset and our developed material taxonomy, either fibre taxonomy or fabric taxonomy; others refer to the labels are not in any taxonomy. TextileNet has rich information in fibre and fabric and is designed with material scientists.}

\label{tab:compare-table}
\begin{tabular}{lcccccc}
\hline
\multirow{2}{*}{Dataset} &
  \multirow{2}{*}{Images} &
  \multicolumn{3}{c}{Material-related Labels*} &
  \multirow{2}{*}{\begin{tabular}[c]{@{}c@{}} Taxonomy \\ based\end{tabular}} \\ \cline{3-5}
             &           & fibre & fabric & others &   &                       \\ \hline
DeepFashion \cite{liu2016deepfashion}  & 800K      & 2     & 16     & 200    & \xmark  \\ 

\textcolor{LightGreen}{FashionAI} \cite{zou2019fashionai}  & 357K       & \multicolumn{3}{c}{dataset not fully disclosed}           & \xmark \\ 
iMaterialist \cite{guo2019imaterialist} & 1M        & 11     & 21     & 2     & \xmark \\ 
CLOTH3D \cite{madadi2021cloth3d++}      & 11.3K     & 2     & 2      & -      & \xmark \\ 
FabricID \cite{wang2018fabric}   & $\sim$20k      & -     & coded name      & -      & \xmark \\
Fabric 1.0 \cite{shen2019large}   & $\sim$46k      & -     & not disclosed      & -      & \xmark \\
TILDA \cite{chetverikov2002finding}        & 3.2K      & -     & 8      & -      & \xmark \\ 
\textbf{TextileNet} (ours)   & 761k  & 33    & 27     & -      & \cmark    \\ \hline
\end{tabular}
\end{table}

\Cref{tab:compare-table} summarises current datasets in the fashion and material domain, extended with our proposed TextileNet dataset. 
TextileNet contains a large number of images (only 4 other datasets in \Cref{tab:compare-table} have more than 500K images), and has detailed fibre and fabric labels (more than 50 labels combined) designed in close collaboration with experts in the material science domain.

Unique to our dataset is that all labels are from the textile taxonomies we developed, and incorporate information on the material source (fibre) and links fibres to fabrics that are the basis for fashion design, that consumers can choose from. 
In this paper, we will demonstrate how this dataset and existing ML models can identify the material origins of user garments. \textbf{TextileNet is connecting the various elements in the production flow to benefit the fashion industry in the identification of materials, reduce fashion waste, and contribute to more sustainable fashion}.

\section{Taxonomy-based approach to textiles}
\label{sec:textile}

Originally, the term ``textiles" referred to woven or knitted products; however, textiles is now an umbrella term to describe textile-related products, including materials, fibres, fabrics, \textit{etc} \cite{kadolph2016textiles,clark2011handbook,grishanov2011structure}. 
\textcolor{LightGreen}{Generally, fibre is defined as a small threadlike structure which is filament or staple \cite{sinclair2015understanding,textileexchange2021preferred}, fibres are the foundation for all textile products \cite{sinclair2014textiles,zhang1995fabric}.
Fabric is a cloth produced by knitting, weaving or non-woven bonded (\textit{e.g.} felting) fibres \cite{zhang1995fabric,sinclair2014textiles}.} 
The taxonomies we developed of textiles, as a basis for our TextileNet dataset, are sub-divided into a fibre taxonomy and fabric taxonomy. The fibre and fabric taxonomies utilise the textiles classification depicted in  \Cref{fig:textile}.

\subsection{Fibre taxonomy}

Fibres are categorized as \textit{natural} or \textit{man-made/artificial} \cite{hearle2008physical,grishanov2011structure,byrne2000technical}. Natural fibres are the fibres derived from natural habitats, they exist in the form of their raw states, such as cotton fibre from cotton plants or wool from animals \cite{kozlowski2012handbook,kozlowski2020handbook,Rowell2008,sinclair2014textiles}. Man-made or artificial fibres are created using industrial methods \cite{grishanov2011structure,horrocks2000handbook}. 

\begin{figure}[!ht]
\centering
  \centering
  \includegraphics[scale=0.4]{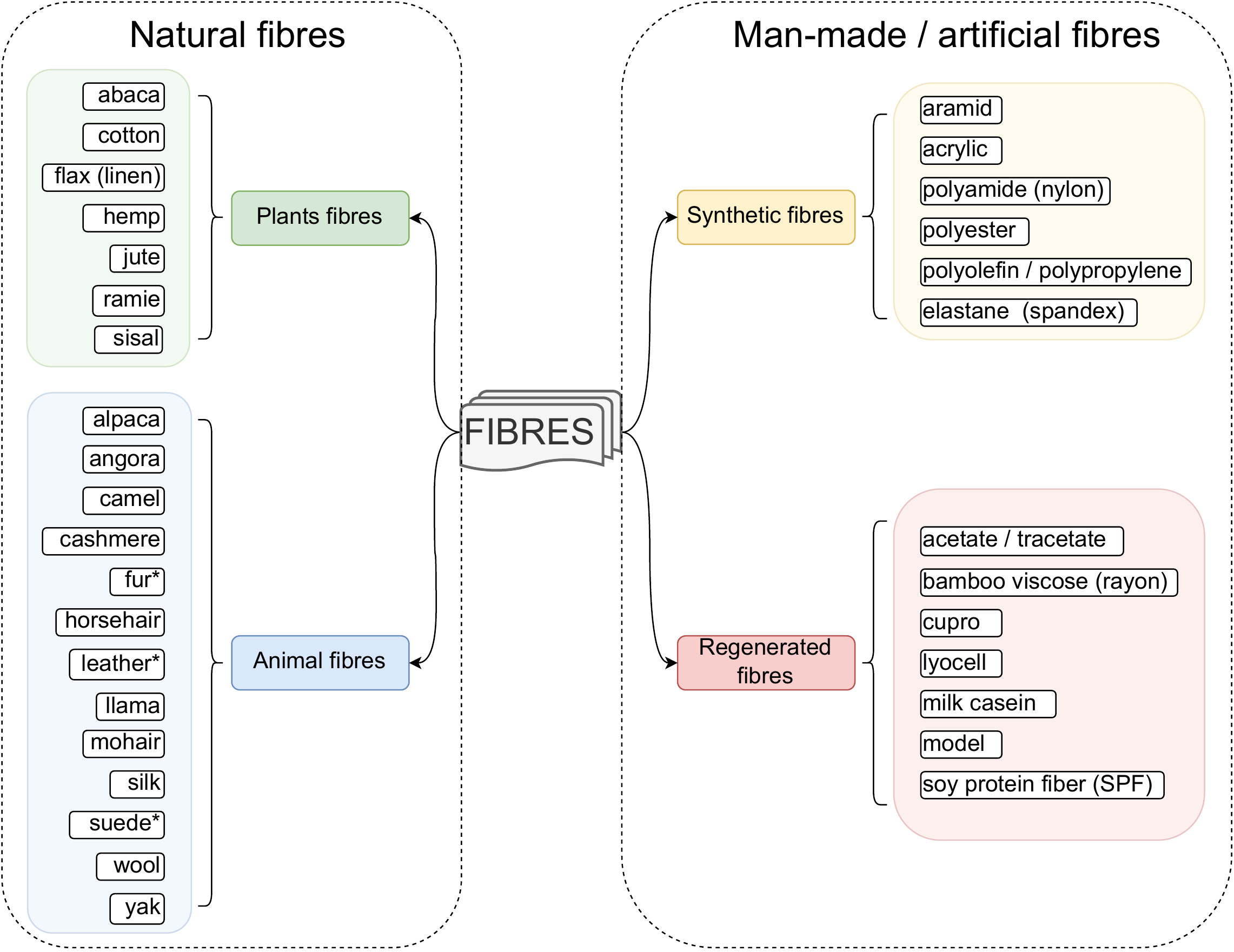}
  \captionof{figure}{Taxonomy for fibres, classified by its fibre category (the four macro-types): Natural fibres (animal and plant), Man-made/artificial fibres (Synthetic fibres and Regenerated fibres). The taxonomy then shows how each fibre can be categorized into these four different macro-types. * are "non-textile" .}
  \label{fig:fibre}
  \vspace{-10pt}
\end{figure}

\Cref{fig:fibre} shows the fibre taxonomy with fibres in the following four main categories (macro-types):
\begin{itemize}
    \item \textit{Plant fibres} are natural fibres that are composed of cellulose and used in their natural fibre shapes, such as cotton and flax \cite{sinclair2015understanding}. 
    \item \textit{Animal fibres}, similar to plant fibre, are natural fibres. Fur, leather, and suede are special cases \textcolor{LightGreen}{classified as natural fibres in this category, they are outliers in the general definition of fibres \cite{eu2011regulationa, sinclair2015understanding,sinclair2014textiles}. We further discuss this in the supplementary material. }
    \item \textit{Synthetic fibres} are formed from petroleum, at the same time, they can also be derived from waste (discarded PET bottles) \cite{mcintyre2005synthetic}.  Polyester, acrylic, polyamide (nylon, aramid), polyolefin, and elastane are examples of synthetic fibres in the textile industry \cite{mcintyre2005synthetic,sinclair2014textiles}. 
    \item \textit{Regenerated fibres} can be obtained from a variety of raw sources, including plants (wood pulp, bamboo, soybeans), milk, eggs and waste (discarded cotton clothing) \textit{etc.} \cite{haule2016preparation,  rex2019possible}. \textcolor{LightGreen}{
    The difference between regenerated fibres and natural fibres is that regenerated fibres are reconstituted from the above-mentioned raw materials and manufactured to shape into new fibres rather than being used in their original form \cite{SINGH2022113,woodings2001regenerated}. Based on the raw sources, they can be named \textit{regenerated cellulose fibres} and \textit{regenerated protein fibres} respectively.} We further include this information in our supplementary material. 
\end{itemize}

The design of the taxonomy with those four macro-types allows future extensions and integration of new fibre categories by other researchers and users of our TextileNet dataset. 

\subsection{Fabric taxonomy}


With regards to the fabric taxonomy, we closely collaborated with material science domain experts as well as drew on potential raw material sources (fibre) \cite{byrne2000technical,Rowell2008,mcintyre2005synthetic,kozlowski2012handbook,kozlowski2020handbook, hearle2008physical,clark2011handbook,bunsell2009handbook} to select the fabrics to be covered in our taxonomy. Since fabrics are made of fibres, we can extend the taxonomy shown in \Cref{fig:fibre}. Due to space limitations, we present the full fabric taxonomy in our Appendix. It is worth noting that it is possible for a fabric to be made of multiple fibres, and we characterise this complex relationship in our fabric taxonomy. The fabric taxonomy also contains three meta-types, and those are \textit{woven}, \textit{knitted} and \textit{non-woven} \cite{horrocks2000handbook}. 

\begin{figure}[!ht]
\centering
  \centering
  \includegraphics[scale=0.45]{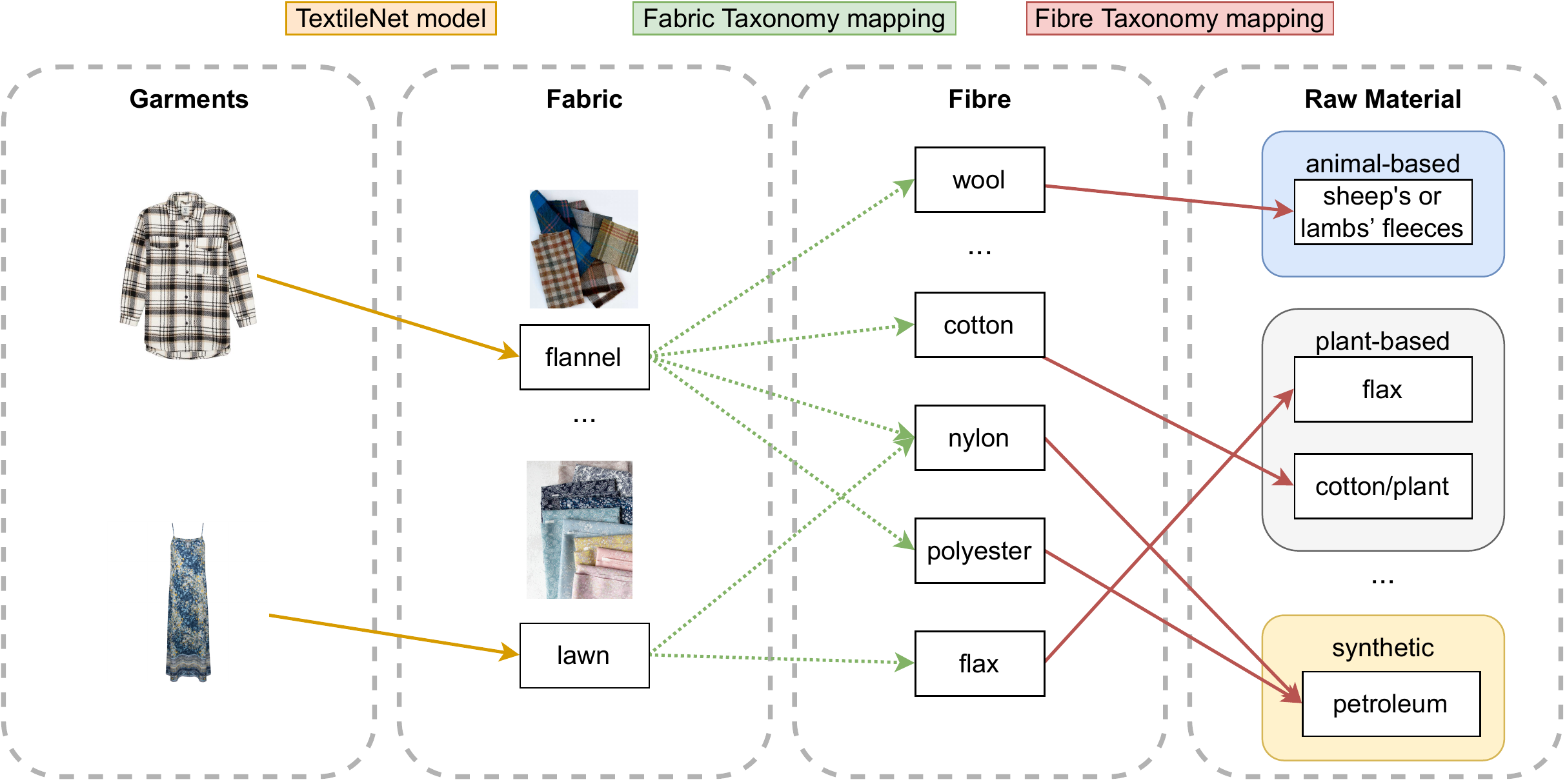}
  \captionof{figure}{We present two garment examples in TextileNet-fabric to predict fabric labels (\textit{flannel} and \textit{lawn}), and show how they can use our taxonomies to create relations to different fibres and raw materials. For instance, for \textit{lawn}, we can connect it to \textit{nylon} and \textit{flax}, this then further connects us to the raw material space with \textit{flax} and \textit{petroleum}.}
  \label{fig:fabric}
\end{figure}

With the designed fibre and fabric taxonomies, one can now trace the origin and production flow of a particular fabric. To further illustrate how the fabric taxonomy and TextileNet-fabric could work in conjunction with the fibre taxonomy, in \Cref{fig:fabric}, we show the first use model to predict the fabric of garments which are \textit{flannel} and \textit{lawn}, and how to utilise the relational information in our taxonomies to build their manufacture flow (fabric taxonomy mapping is in green dash and fibre taxonomy is in red line). For example, for \textit{lawn}, we can connect it to \textit{nylon} and \textit{flax}, this then further connects us to the raw material space with \textit{flax} and \textit{petroleum}.
It is worth noting that this trace back process not only characterises the manufacturing process but also identifies the raw material categories (\textit{animal-based or synthetic}). From the production point of view, these taxonomies would have the potential to help researchers or designers to estimate the life-cycle-analysis researchers can then assign \cite{macarthur2013towards} to garments. For instance, we can assign a certain carbon emission number to \textit{lawn} $\to$ \textit{flax} and \textit{lawn} $\to$ \textit{nylon} to model the transition between fibre to fabric. The carbon footprint can also be estimated based on these raw materials on the right of \Cref{fig:fabric}. These taxonomies then help materialise the idea of creating a textile circular economy discussed in material science \cite{laitala2012sustainable, macarthur2013towards}.

\section{The TextileNet dataset}

The TextileNet dataset consists of two parts, TextileNet-fiber and TextileNet-fabric, which were developed based on two taxonomies - fibre taxonomy and fabric taxonomy, respectively.
Future users can also extend the fabric or fibre taxonomy for broader range of applications, for instance, an interactive design for customers will help them understand what is the garment made of and with the development of carbon footprint calculation, customers can raise awareness of sustainable fashion.

\textbf{Taxonomy-informed image collection}
We construct our dataset by combining Google Images (41.65\%) with existing fashion datasets (58.35\%) containing labels in our taxonomies. We have taxonomy-based labels first, and we collect images using query phrases derived from our taxonomies, as opposed to previous fashion datasets that crawl metadata and explore information of collected metadata to design labels.
\begin{figure}[h]
\centering
  \centering
  \includegraphics[scale=0.3]{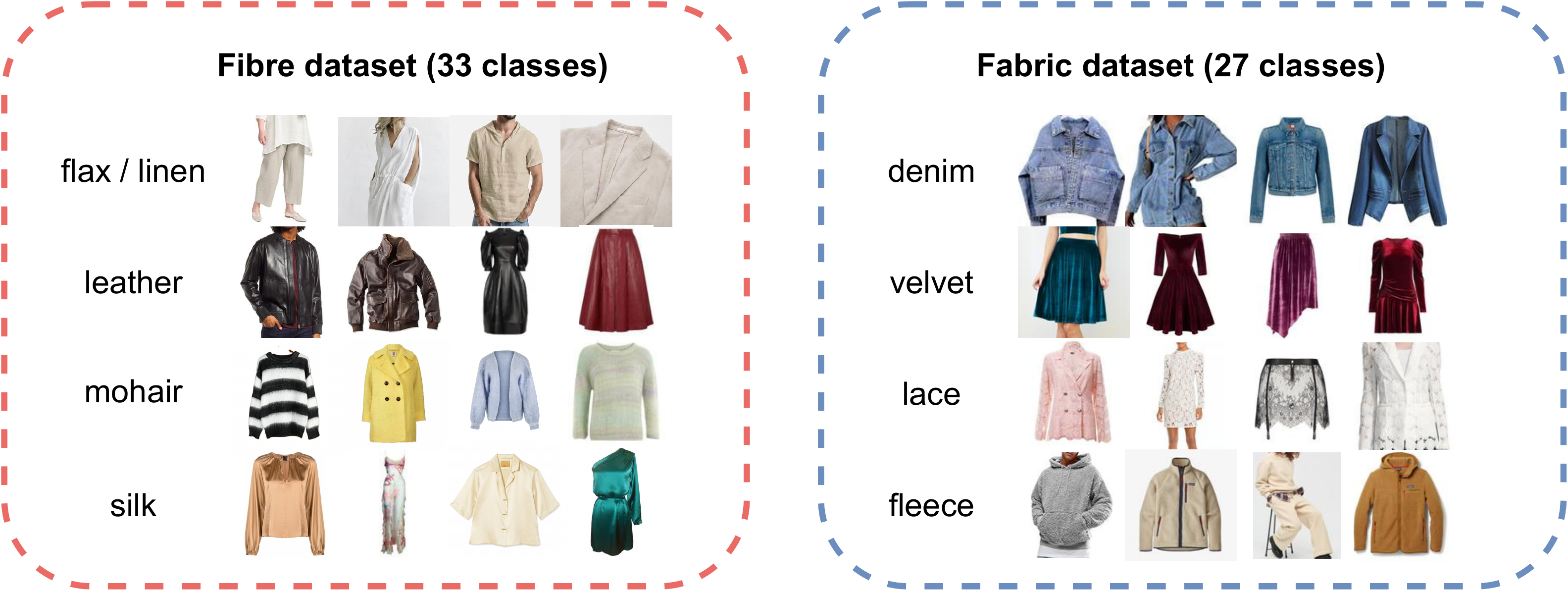}
  \captionof{figure}{Samples of fibre dataset (left) and fabric dataset (rights). This is a visualization to show what data is included in TextileNet, these are real images sampled from TextileNet.}
  \label{fig:sample}
  \vspace{-10pt}
\end{figure}

\textbf{Google images}
We collected images from Google's Image Search. In practice, we found simply searching for the fabric or fibre names would not return meaningful results. Our solution was to bound this search to obtain garment items rather than textile items. We used labels from our fabric/fibre taxonomy as keywords and combined them with clothing categories to build a query to the Image Search, \textit{i.e.} ``fibre labels + category". For instance, for a clothing category such as ``cardigan", we can combine it with a fabric/fashion label keyword to build the search phrase ``wool cardigan".  We use the fashion clothing category labels defined by Zheng \textit{et al.}, since these labels are widely utilised in fashion datasets \cite{zheng/2018acmmm}.


\textbf{Reconstruct fashion datasets}
As there are existing datasets with partial material labels, we reconstructed two fashion datasets with material labels to fit into our dataset, \textit{i.e.,} iMaterialist  \cite{guo2019imaterialist} and Amazon review data \cite{ni2019justifying} based on our taxonomy-based labelling.
We removed images with multiple human faces and cropped images with human faces. This process was applied to ensure that no human faces/identifiable information are included in our developed dataset. \Cref{fig:sample} shows the samples from the fibre dataset and fabric dataset. 
Our final fibre and fabric datasets have 33 and 27 classes respectively.




\section{Dataset validation and baselines}
This section provides a quantitative evaluation. In \Cref{sec:setup} we describe our evaluation setup and \Cref{sec:results} reports the accuracy of applying a CNN and a ViT model on our dataset.

\textbf{Experiment setup}
\label{sec:setup}
We validated our dataset with popular Deep Learning models to establish a baseline comparison. 
We considered the following two models:
\begin{itemize}
    \item \textcolor{black}{\textit{ResNet50}} \cite{he2016deep}: The ResNet family networks make use of the residual connections and we use the ResNet50 structure as a representative Convolutional Neural Network. 
    \item \textit{Vision Transformer (ViT)} \cite{dosovitskiy2020image}: There is now a trend for Transformer models to become the de-facto standard architecture for various domains of tasks. We provide a ViT as a baseline model for our proposed TextileNet dataset.
\end{itemize}

For both the \textcolor{black}{ResNet50} and the ViT model (tiny), we use their standard architecture setup for training on the TextileNet-fibre and TextileNet-fabric in our TextileNet.

We used the ADAM optimizer \cite{kingma2014adam} for these datasets, and set the learning rate to $5e^{-4}$ and batch size to $256$. We train for 70 epochs on these datasets. We partition our data to 0.7, 0.1 and 0.2 for training, validation and testing. The validation images are randomly sampled from our previous training set in the paper that had the 0.8 and 0.2 training and testing splitting. In our supplementary material, we provide the image process, detailed hyperparameters and more results when training models using $80\%$ to train and $20\%$ for testing.
We run each training 3 times with different random seeds and report their final test accuracy with mean and standard deviations. 
All experiments complete in $<14$ GPU-days on a four NVIDIA GeForce GTX 1080 Ti system with an Intel(R) Xeon(R) CPU E5-2620 v4 at 2.10 GHz.

\begin{table}[h]
\centering
\caption{Top-1, top-3 and top-5 accuracy on TextileNet-fibre and TextileNet-fabric with a validation set.}
\label{tab:results}
\begin{tabular}{@{}c|ccc|ccc@{}}
\toprule
     & \multicolumn{3}{c}{TextileNet-fibre} & \multicolumn{3}{c}{TextileNet-fabric} \\  
\midrule
Metric        & top-1 & top-3 & top-5 & top-1 & top-3 & top-5 \\ \midrule 
\textcolor{black}{ResNet50} & \textcolor{black}{$49.74 \pm $0.27} & \textcolor{black}{$76.04 \pm $0.13} & \textcolor{black}{$87.32 \pm $0.14} & \textcolor{black}{$65.28 \pm  $0.67} & \textcolor{black}{$85.06 \pm $0.43} & \textcolor{black}{$90.36 \pm $0.31} \\
ViTs     & $47.04 \pm $0.33 & $64.86 \pm $0.10 & $80.04 \pm $0.07      & $61.62 \pm $0.09 & $78.78 \pm $0.11 & $89.20 \pm $0.17 \\
\bottomrule
\end{tabular}
\end{table}



\textbf{Results}
\label{sec:results}
\Cref{tab:results} summarizes the performance of different ML models on classification of different partitions of TextileNet. It is observed that the accuracy on Fabric is generally greater than the accuracy on Fibre. When we look at only the top-5 accuracy, both ML models show relatively good performance (above $80\%$).

Our results in \Cref{tab:compare-table}
also suggests that it is more challenging to recognise fibre than fabric. \Cref{tab:compare-table} shows that the top-1 accuracy of fibre classification is around $15\%$ lower compared to fabric model when we use a ResNet50 model, the same trend can be seen on the ViT model (the top-1 accuracy gap is around $14\%$). Intuitively, classifying fibre should naturally be a harder task, because fibre is closer to the start of the supply chain compared to fabric, as illustrated in \Cref{fig:textile}. This then makes the recognition of fibre a lot more challenging than fabric. \textcolor{black}{
The results also vary from different models, we assume that convolutions might be better at finding texture-related attributes and attention mechanisms might be less powerful on this since they focus on patched sequences. } 

We release these pre-trained models with our dataset so that future users can have direct access to our baseline ML models.


\section{Discussion, broader impact and future work}
Here, we address the future work that can be derived from this dataset, as well as its limits.

\textbf{Standardizing textile materials labelling in datasets through taxonomies}
In the booming textile industry, there is always a drive to discover new fibres and fabrics that are maybe more cost-effective or more environmentally friendly. There is always a worry that fashion datasets published today will soon be obsolete because fashion items get updated quickly. 

Our solution to this problem is to build TextileNet, a dataset that is based on fibre and fabric taxonomies collaboratively designed with material scientists, connecting the different elements in the production flow, starting from the source material. These taxonomies allow future extension with new fibre and fabric types using a taxonomy-informed image-based textile materials retrieval approach. 


\textbf{Broader impact}
As discussed in \Cref{sec:related:why}, the fashion industry is actively thinking of a transformation to be more sustainable. With the global drive of Net Zero \cite{macarthurfashion}, it is now a recommended practice to ensure that textile materials of garments can be traced back to their original sources so that the recycling process can be more carefully designed and their carbon emissions can be more precisely estimated. The dataset presented in this paper shows the potential social and environmental impact in helping researchers and the industry to make progress by providing them with a large-scale dataset. 

Our dataset is also the first of its kind that provides an expert-designed taxonomy for image labelling. For the ML community, we bridge the knowledge gap between textile material research in the fashion industry and ML research, so that novel and existing ML techniques can be implemented and evaluated on our dataset to enable or optimise various fashion applications. From the material science perspective, we make sure that the designed taxonomies are extendable and can be used as a foundation taxonomy for new fibres and fabrics.

\textbf{Limitations and future work opportunities}
Although we have the greatest number of fibre labels compared to other available fashion datasets, there are several limitations in our data collection method:
\begin{itemize}
    \item The query phrases for fashion items are limited by category types, not all fashion items are addressed;
    \item There is an imbalance between the number of samples from different labels. This is due to some fibre or fabric types being less popular. Some new fibre types, such as Soybean fibres and Milk casein, have fewer images than other typical fibres (cotton, wool, \textit{etc.}) due to they are not widely used yet;
    \item We have not included fibre blends in our taxonomy or dataset due to the lack of solid metadata for fibre blends. Fibre blends combine two or more fibres into a single fibre strand or yarn. Fibre blends are often developed to provide a particular level of comfort or reduce the costs of fibres \cite{ribul2021mechanical}. Many blends exist such as wool-acrylic blends for knitted garments, or cotton-polyester in different types of garments \cite{dissanayake2021fabric}. At the moment, it is challenging to identify and recycle them \cite{haslinger2019upcycling}. Future work can integrate fibre blends into the fibre taxonomy.
    \item  Label noise is inescapable in large-scale datasets generated by online queries and crowdsourcing \cite{wei2021open}, such as the one we implemented here. This is another limitation imposed on our work. Due to the absence of blend fibres in the taxonomy, certain images labelled with two fibres or one fibre and one fabric may also be regarded to have closed-set noisy labels. A garment that was labelled as both wool and acrylic in our dataset is an example of closed-set noisy labelling, despite being made of a wool-acrylic blend containing both wool and acrylic. It is a missing piece in the industry and material research that textile material metadata lacks clear blend information for textiles online where it may only be found on the wash label on the garments and is not constant. 
\end{itemize}

\textcolor{LightGreen}{
Despite these challenges, we expect our dataset can provide a standardization on labelling image data in the fashion industry. One potential drawback of the dataset is that its contained images are from online images and this inevitably contains label noises with varying image resolutions. The preferred method for upstream supply chain companies is to provide high quality images of their products.}
Companies or organizations in the supply chain can also use their fetched high-quality images to construct their own datasets based on the fibre and fabric taxonomies provided in this paper.

\textbf{Future work}
We expect that this work will contribute to the development of a new fashion AI-based platform in the future. A textile recognition technique can be created based on this work. The TextileNet is mainly built for image-based textile material retrieval and contributes to the textile circularity process, including the supply chain and the consumer experience. 

\textcolor{LightGreen}{
We anticipate that the implementation of a taxonomy-based label can provide a standard for dataset and encourage material scientists to reflect on this labelling problem on fibre blends. Future work can investigate the image quality through image-wise manual inspection on the entire dataset.}

For the textile supply chain, it is possible to develop advanced sorting techniques based on the taxonomy and TextileNet dataset. The TextileNet-fibre has the greatest number of fibre labels among existing fashion datasets, allowing it to meet the fundamental requirements of textile material classification. The designed taxonomies would also help identification of the raw materials and production methods for different textiles. This can be used in a coarse-grained classification before using professional spectrum-based analysis for material composition classification. The goal is to scale up recycling and optimize the textile circularity supply chain. As for consumers, they can take a picture of the garments and learn about the fabric and fibres that are likely used to make the garment. They can then use this to better understand whether the garment is made of natural fibres or synthetic fibres, what the raw material is and maybe learn how it can be recycled.

\section{Conclusion}
In this paper, we introduced the TextileNet dataset. This dataset is designed based on a fibre and a fabric taxonomy created with material domain experts. We demonstrate the definition of fibres, fabrics and textiles to better understand the use of the dataset. The TextileNet-fibre contains 33 labels and the TextileNet-fabric part has 27 labels. These two parts combined have around 760K images. 

TextileNet-fibre can be used to trace raw material by fibre taxonomy, and we presented a piece of fabric that can be traced back to its origins and manufacturing processes using our taxonomies. We then discussed how this image-based textile material retrial can be a crucial step to help the fashion industry to meet its sustainability goal. We present and report baseline models (CNN and ViT respectively) on TextileNet, and will open source both the dataset and these pre-trained models.


\section*{Acknowledgments}
This work was supported by the UK Research and Innovation (UKRI) National Interdisciplinary Circular Economy Centres Research programme, as part of the Textiles Circularity Centre (TCC) [grant number EP/V011766/1]. For the purpose of open access, the author has applied a Creative Commons Attribution (CC BY) licence to any Author Accepted Manuscript version arising.

\newpage

\bibliographystyle{abbrv}
\bibliography{references}
\newpage


\appendix

\section{Appendix on Fibre taxonomy}

This section presents the fibre taxonomy in \Cref{tab:fibre-tax}. 
The taxonomy for fibres is classified by its fibre categories (the four macro-types): animal, plant, synthetic and regenerated. There are 442,035 images in TextileNet-fibre, the number of images for the 33 fibre labels is shown in \Cref{fig:fibre}. For easy of overview, we use log in the y-axis and added the actual numbers.


\begin{table}[h]
\adjustbox{max width=\textwidth}
\centering
\caption{Fibre taxonomy, raw materials are listed. \textcolor{LightGreen}{The first two parts are Natural fibres (plant fibre refers to natural plant fibre), followed by Regenerated fibres and Synthetic fibres.}}
\label{tab:fibre-tax}
\begin{tabular}{@{}cll@{}}
\toprule
\textbf{Raw material}         & \textbf{Fibre name}         & \textbf{Type of fibre}      \\ \midrule
cotton                        & cotton                      & plant fibre \\
abaca                         & abaca                       & plant fibre      \\
flax                          & linen/flax                 & plant fibre       \\
hemp                          & hemp                        & plant fibre       \\
jute                          & jute                        & plant fibre       \\
ramie                         & ramie                       & plant fibre       \\
sisal                         & sisal                       & plant fibre      \\ \midrule
sheep's or lambs’ fleeces     & wool                        & animal fibre       \\
alpaca                        & alpaca                      & animal fibre        \\
angora rabbit                 & angora                      & animal fibre        \\
camel                         & camel                       & animal fibre        \\
goat                          & cashmere                    & animal fibre        \\
horse                         & horse hair                  & animal fibre        \\
llama                         & llama                       & animal fibre        \\
angora goat                   & mohair                      & animal fibre        \\
insects                       & silk                        & animal fibre        \\
yak                           & yak                         & animal fibre        \\
animal skin                   & leather* \footnotemark{}                    & animal non-textile  fibre           \\
animal skin                   & suede*                       & animal non-textile fibre           \\
animal fur                    & fur*                         & animal non-textile fibre           \\ \midrule
bamboo, wood                        & viscose/rayon      & regenerated cellulose fibre \\
\multirow{4}{*}{wood, plants} & acetate/triacetate        & regenerated cellulose fibre \\
                              & cupro                       & regenerated cellulose fibre \\
                              & modal                       & regenerated cellulose fibre \\
                              & lyocell                     & regenerated cellulose fibre \\
milk fibre                    & milk fibre/milk casein    & regenerated protein fibre   \\
soybeans                      & soybean protein fibre (SPF) & regenerated protein fibre   \\ \midrule
petroleum                     & polyester                   & synthetic fibre              \\
petroleum                     & aramid                      & synthetic fibre              \\
petroleum                     & acrylic                     & synthetic fibre              \\
petroleum                     & polyamide(nylon)            & synthetic fibre              \\
petroleum                     & polyolefin/polypropylene  & synthetic fibre              \\
petroleum                     & elastane/spandex          & synthetic fibre              \\ \bottomrule
\end{tabular}
\end{table}

\footnotetext[1]{The fibres with * are the special cases fall outside the standard definition for fibre. According to Sinclair \textit{et al.} \cite{sinclair2015understanding} and EU Regulation 1007/2011 \cite{eu2011regulationa}, they are “non-textile animal origin” and special cases in natural fibre, we here classified them in animal fibre as animal non-textile fibre}

\subsection{}
\begin{figure}
\centering
  \centering
  \includegraphics[scale=0.7]{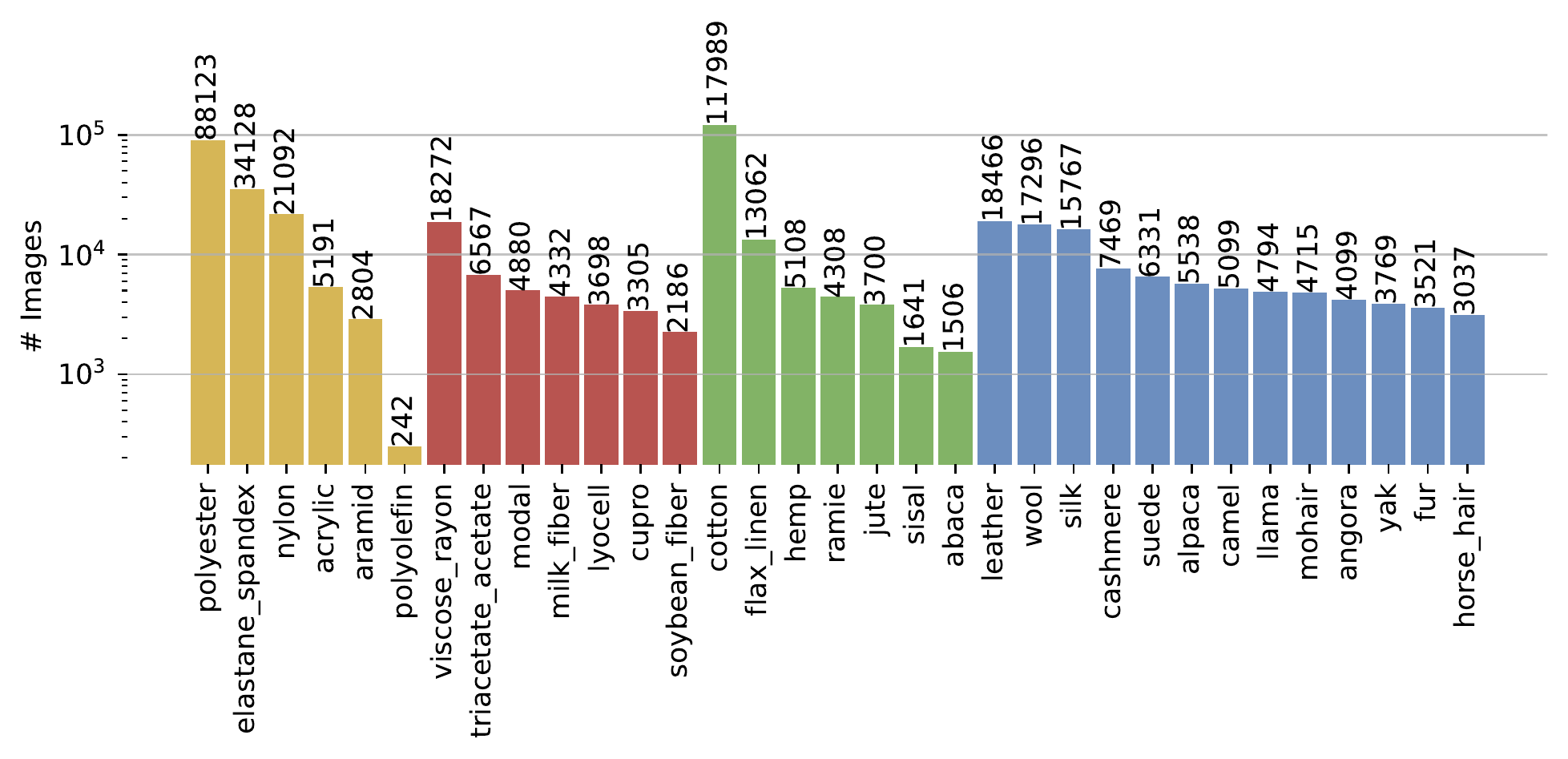}
  \captionof{figure}{\textcolor{LightGreen}{Image number of fibre labels in each macro type: synthetic, regenerated, plant and animal.}}
  \label{fig:fibre_chart}
\end{figure}

Future scientists with new fibre types can add to \Cref{tab:fibre-tax}, but they would also need the raw material information to be able to complete the taxonomy.

\section{Fabric taxonomy}

\begin{figure}[!h]
\centering
  \centering
  \includegraphics[scale=0.7]{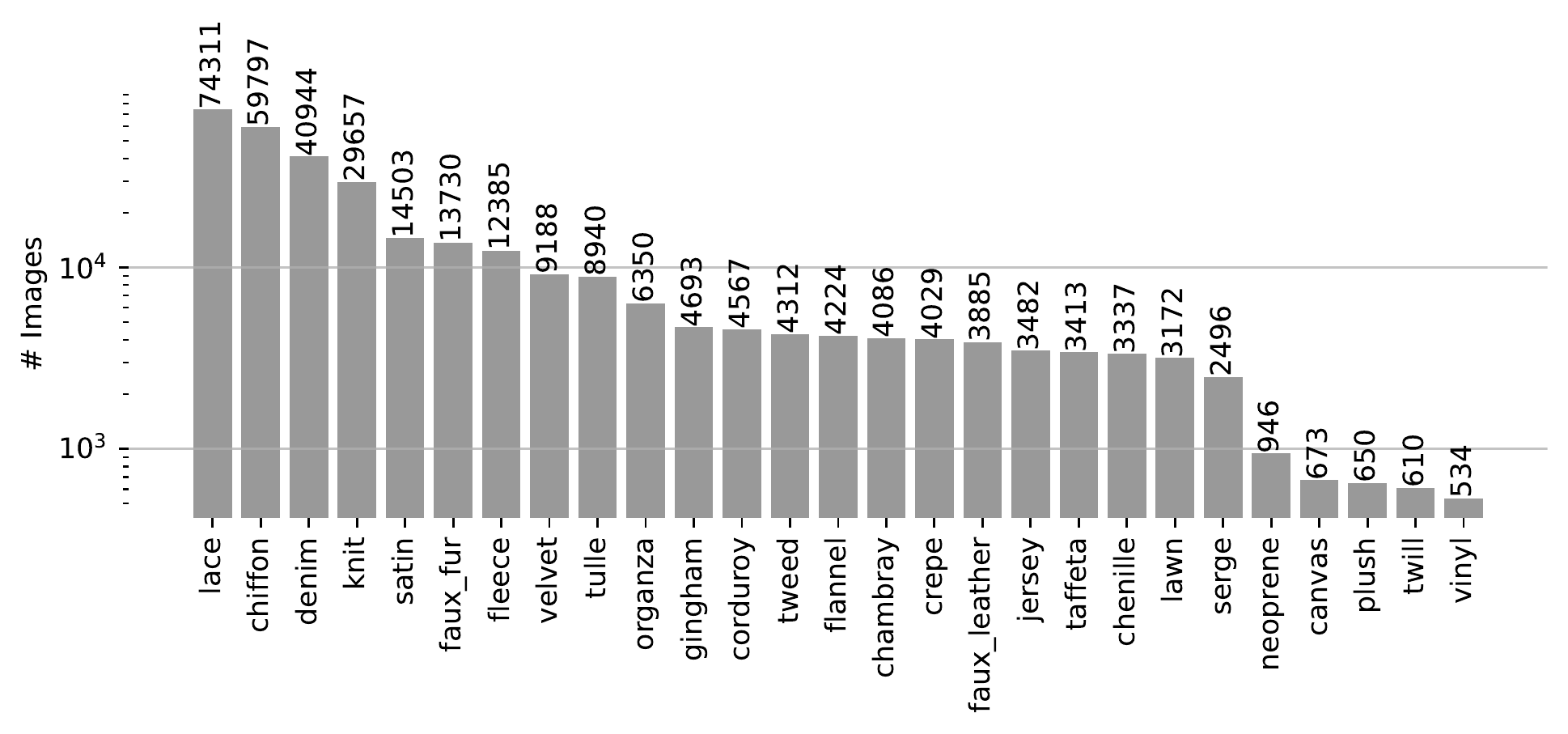}
  \captionof{figure}{\textcolor{LightGreen}{Image number of fabric labels.}}
  \label{fig:fabrics}
\end{figure}

Similar to the previous section,
here we have the common clothing fabrics. 
Fabric can be made from different fibres and there is normally a production method involved when making fabrics from fibres.
The Fabric taxonomy in \Cref{tab:fabric-tax} displays this mapping of the fibre-to-fabric production process. There are 318,914 images in TextileNet-fabric, the number of images for the 27 fabrics labels is shown in \Cref{fig:fabrics}.

It is worth noting the mapping between fibre and fabric can aid in tracing fabric's origin, manufacturing method and what are possible raw materials. The fabric names in  \Cref{tab:fabric-tax} are the TextileNte-fabric labels. 
\begin{table}[h]
\centering
\caption{Fabric taxonomy, each fabric is listed with common fibres making it. \textcolor{LightGreen}{* are non-woven fabric, they are formed directly by chemically bonded polymer (plastic).}}
\label{tab:fabric-tax}
\begin{tabular}{@{}ll@{}}
\toprule
\textbf{Fabric name} & \textbf{Common fibres/material for fabric}                          \\ \midrule
canvas               & cotton, hemp, flax                                         \\
chambray             & cotton, flax                                               \\
chenille             & wool, cotton, silk, rayon                                  \\
chiffon              & silk, cotton, nylon, polyester                             \\
corduroy             & cotton, wool, silk, polyester, rayon                       \\
crepe                & silk, wool, synthetic                                      \\
denim                & cotton, poly cotton                                        \\
faux fur             & acrylic, silk, wool, mohair, cotton, polypropylene(PP)     \\
faux leather* \footnotemark          & PU, PVC, polyvinyl                                         \\
flannel              & wool, cotton, synthetic fibres                             \\
fleece               & polyester                                                  \\
gingham              & cotton                                                     \\
jersey               & wool, silk, cotton, synthetic fibres                       \\
knit                 & silk, linen, cotton, wool, viscose, rayon                  \\
lace                 & nylon, polyester, rayon, cotton, silk, flax                \\
lawn                 & flax, cotton                                               \\
neoprene             & polychloroprene, synthetic rubber                          \\
organza              & silk, nylon, rayon, polyester                              \\
plush                & polyester, wool, cotton, viscose, silk                     \\
satin                & silk, wool, nylon, polyester                               \\
serge                & wool, silk                                                 \\
taffeta              & silk, nylon, acetate, rayon, polyester                     \\
tulle                & silk, cotton, polyester, nylon, rayon                      \\
tweed                & wool, silk, rayon                                          \\
twill                & cotton, polyester                                          \\
velvet               & silk, cotton, flax, wool, rayon, polyester, nylon, acetate \\
vinyl*                & polyvinyl                                                  \\ \bottomrule
\end{tabular}
\end{table}

\footnotetext{Fabric name with * are non-woven fabric, they are formed directly by chemically bonded polymer (plastic)}


\section{Dataset construction and cleaning}
As we introduced in the paper, our dataset combines existing datasets and results from Google Image Search.
\subsection{Images from Google Images}
\paragraph{Collecting Images from Google Images}

\begin{table}[h!]
\centering
\caption{Garment categories for query phrases}
\label{tab:categories}
\begin{tabular}{@{}llllllll@{}}
\toprule
\multicolumn{8}{c}{\textbf{Category names}}                           \\ \midrule
blazer               & blouse & cardigan      &dress        &hoodie   &jumpsuit    & pants   &shirt                                      \\
shorts &skirt         &sweater  &vest &outerwear      & sweatshirt       &t-shirt    &top                         \\ \bottomrule
\end{tabular}
\end{table}
 We used labels from our fibre and fabric taxonomies as keywords and combined them with clothing categories to build query phrases. There are 16 garment categories shown in \Cref{tab:categories}, they are commonly used in fashion datasets \cite{zheng/2018acmmm} and we used them to combine our label from taxonomy to build the query phrases. They are ``blazer, top, t-shirt, dress, hoodie, shirt, blouse, cardigan, pants, skirt, sweater, sweatshirt, shorts, vest, outerwear, jumpsuits". \textcolor{LightGreen}{Initially, we had 36 fibre labels and 30 fabric labels. This process results in a list of 576 fibre unique quires and a list of 480 fabric unique quires.} Then we feed these queries to the Google Image Search with the usage right under \textit{labelled-for-reuse-with-modifications}. The images are downloaded with their metadata including image descriptions, image filenames and URLs.\textcolor{LightGreen}{ A total of 359,078 images are collected from Google Images, which takes 41.65\% of the dataset.}

\paragraph{Data Cleaning}
We cross-checked the downloaded image descriptions with query phrases, and removed images without any label information in their descriptions. \textcolor{LightGreen}{After the removal based on metadata, We analysed the distributions of images returned from the queries. We then used random sampling and manually inspected for images. During the process, we removed following labels and images:}
\begin{itemize}
    
    \item \textcolor{LightGreen}{We excluded \textit{coir}, \textit{kapok} and \textit{Azlon} in fibre taxonomy. Both \textit{coir} and \textit{kapok} are natural plant fibres, but they are commonly used in textile furniture such as matting and bedding. New sustainable regenerated fibre \textit{Azlon} is removed because neither the quantity nor quality of images is sufficient for the dataset.}
    \item We removed the query which has less than 100 images;
    \item \textcolor{LightGreen}{We randomly sampled query phrases for each fibre/ fabric label and manually went through the images with their associated labels;}
    \item \textcolor{LightGreen}{We removed unusable images that are of low resolution, incorrect for the label, multi objects or whose dominant objects are irrelevant to clothes.}
\end{itemize}

\textcolor{LightGreen}{We then randomly sampled a subset (around 0.1\%) of the dataset and manually went through the dataset with its associated labels. In this small subset, we suggest around 82\% of images are matching with our labels.}

The images are then sorted by fibre and fabric labels described in the two taxonomies in \Cref{tab:fibre-tax} and \Cref{tab:fabric-tax}.

\subsection{Collection of other fashion datasets}

We combined and reconstructed based on two fashion datasets, iMaterialist  \cite{guo2019imaterialist} (Creative Commons Attribution 4.0 License) and Amazon review data \cite{ni2019justifying} respectively, and imported some of their images into our dataset. \textcolor{LightGreen}{The number of images from iMaterialist is 480,558 images and 22,588 from Amazon Review, takes 55.73\% and 2.62\% of the dataset respectively.}
When incorporating images in these datasets, we use their material attributes to assign images to our taxonomy-based labels. Images with attributes that do not match any of our taxonomy-based labels are then ignored.

The iMaterialist dataset contains labels for materials, and some images have multiple labels. To prevent this from causing confusion and adding noise to our data, we remove images that have more than two material labels in the same taxonomy. 
The Amazon review dataset, on the other hand, is more complex. This dataset contains information of items on the Amazon site and also user reviews of these items. Obviously, the information of these fashion products would contain images. We then use the similar query phrase approach described in the previous subsection to collect data. We use fibre/fabric label + category as a query phrase, and use this query phrase to traverse the fashion part of the dataset (The Amazon dataset has several parts and the fashion part contains all the items in their fashion department).

After combining all sources and giving them correct labels based on our taxonomies, we utilised a face detection model from \url{https://github.com/ageitgey/face\_recognition/} to detect whether these images contain human faces. This function returns the number of faces and the top, right, bottom, and left coordinates of each face (in pixels). We removed images with several faces and cropped images with a single human face from the bottom coordinates. We then similarly relied on body detection to remove images without plausible garments and when multiple garments are existing.

\section{Training details}
\textcolor{LightGreen}{\paragraph{Data processing} We resized the image to 256*256, then applied randomised cropping to size 224*224, this is then followed by a randomised horizontal flip. At evaluation time, all images are centralised and resized to size 224*224.}

\paragraph{Baseline}

The models that we evaluated on the proposed TextileNet dataset are trained from scratch, these models are ResNet18 \cite{he2016deep} and Vision Transformer (Vit) \cite{dosovitskiy2020image}. For both the ResNet18 and the ViT model, we use their standard architecture setup, \Cref{resnet} and \Cref{vits} show the parameters, respectively.

\begin{table}[h]
  \caption{ResNet18 parameters \\ (* stands for the number of classes)}
  \label{resnet}
  \centering
  \begin{tabular}{ccccc}
    \toprule
    Layer Name     & Output Size     & kernel size  & channels  &stride  \\
    \midrule
    conv1      & $112 \times 112 \times 64$  & $7 \times 7$ & 64 & 2    \\
    \midrule
    maxpool        &   & $3 \times 3$ & 64  & 2     \\
    \midrule
    block\_0\_conv0\_x      & \multirow{2}{*}{$56 \times 56 \times 64$} & $3 \times 3 $ &64 &2   \\
    block\_0\_conv1\_x      & & $3 \times 3$ & 64 & 1 \\
    \midrule
    block\_1\_conv0\_x      & \multirow{2}{*}{$28 \times 28 \times 128$} & $3 \times 3 $ &128 &2   \\
    block\_1\_conv1\_x      & & $3 \times 3$ & 128 & 1\\
    \midrule
    block\_2\_conv0\_x      & \multirow{2}{*}{$14 \times 14 \times 256$} & $3 \times 3 $ &256 &2   \\
    block\_2\_conv1\_x      & & $3 \times 3$ & 256 & 1\\
    \midrule
    block\_3\_conv0\_x      & \multirow{2}{*}{$7 \times 7 \times 512$} & $3 \times 3 $ &512 &2   \\
    block\_3\_conv1\_x      & & $3 \times 3$ & 512 & 1 \\
    \midrule
    average pool     & $1 \times 1 \times 512$       & $7\times7$ average pool  \\
    \midrule
    fully connected & n* &$512 \times n$&\\
    \midrule
    softmax & n \\
    

    \bottomrule
  \end{tabular}
\end{table}

\begin{table}[h]
  \caption{Vision Transformers parameters}
  \label{vits}
  \centering
  \begin{tabular}{cccc}
    \toprule
    Patch size 
    & Embedding dimension
    & Depth
    & Number of heads \\    
    \midrule
    16 & 192 & 12 & 3\\
    \bottomrule
  \end{tabular}
\end{table}

We used the ADAM optimizer \cite{kingma2014adam} for these datasets and set the learning rate to $5e^{-4}$ and batch size to $256$. We train for 70 epochs on these datasets. We use $80\%$ of the data to train and evaluate the rest $20\%$ on the final model in the paper.

\paragraph{More Result}
We run each training 3 times with different random seeds and report their final test accuracy with mean and standard deviations. 

Different from the results presented in our paper, We use $80\%$ of the data to train and evaluate on the rest $20\%$ on the final model. \Cref{tab:results2} demonstrates the accuracy of different baseline models on this dataset setup, it shows a similar trend to the results presented in our paper.


\begin{table}[tbp]
\centering
\caption{Top-1, top-3 and top-5 accuracy on TextileNet-fibre and TextileNet-fabric.}
\label{tab:results2}
\begin{tabular}{@{}c|ccc|ccc@{}}
\toprule
     & \multicolumn{3}{c}{TextileNet-fibre} & \multicolumn{3}{c}{TextileNet-fabric} \\ 
\midrule
Metric        & top-1 & top-3 & top-5 & top-1 & top-3 & top-5 \\ \midrule 
ResNet18 & $49.35 \pm $0.22 & $74.45 \pm $0.17 & $82.93 \pm $0.04 & $58.17 \pm 0.40 $ & $81.14 \pm $0.36 & $88.68 \pm $0.19 \\
ViTs     & $44.39 \pm $0.08      & $69.00 \pm $0.06      & $78.51 \pm $0.04      & $58.51 \pm $0.04 & $80.21 \pm $0.04 & $87.77 \pm $0.07 \\
\bottomrule
\end{tabular}
\end{table}

All experiments complete in $<14$ GPU-days on a four NVIDIA GeForce GTX 1080 Ti system with an Intel(R) Xeon(R) CPU E5-2620 v4 at 2.10 GHz.

\section{Discussion}

Nowadays, a low-cost, high efficiency technique for the automatic identification of textile materials in garments is missing. The TextileNet is mainly built for image-based textile material retrieval and contributes to the textile circularity process. As described in the paper, the recycling process for textile materials relies heavily on raw materials. The TextileNet-fibre has the greatest number of fibre labels among existing fashion datasets, allowing it to meet the fundamental requirements of textile material classification. \textcolor{LightGreen}{We compare the related works in \Cref{tab:compare-fabric}.}

\begin{table}[h!]
\centering
\caption{Compare Image-based material retrieval with traditional methods}
\label{tab:compare-fabric}
\begin{tabular}{lllll} \\ \hline
Device & Cost  & Applications  & Accuracy  & Real-time  \\ \hline
NIR spectrum  \cite{liu2020qualitative} & above \$10,000  & fibre classification  & 96\% \footnotemark[3] & \xmark \\
ATR-FTIR \cite{riba2020circular} & above \$20,000  & fibre classification & 100\% \footnotemark[4]  &  \xmark \\
ATR-FT-IR \cite{peets2017identification} & above \$20,000  & fibre classification & partially \footnotemark[5] &  \xmark \\
Thermal camera  \cite{yildiz2016thermal} & above \$4,300 &fabric defect  \footnotemark[6]  &  96\% & \cmark  \\
RGB camera  & \$30 &TextileNet  & 80\% - 90\% \footnotemark[7] & \cmark  \\ \hline

\end{tabular}
\end{table}

\footnotetext[3]{For only 263 selected samples in total.}
\footnotetext[4]{Only 7 types of fibre classification, all fibres are included in our taxonomy.}
\footnotetext[5]{Only partial silk and wool can be distinguished.}
\footnotetext[6]{Defect detection such as holes, cracks etc. Not applicable for fibre classification.}
\footnotetext[7]{Top5 accuracy for fibres and fabrics.}

Apart from the post-consumed garments as we discussed, there are alternative material tracing solutions for new garments. One option is adopting blockchain for the clothing supply chain. Francisco \textit{et al.} built a conceptual clothing supply chain blockchain model to improve supply chain transparency and thus track the textile footprint \cite{francisco2018supply}. 
The textile industry is trying to integrate blockchain tokens into the supply chain to create a new way to manage carbon footprints \cite{lam2019textile}.

Radio frequency identification (RFID) tags are among the most promising technologies for tracking objects and managing supply chains. However, RFID's use in textiles is still constrained by its high cost, potential security risks, and privacy concerns. \cite{nayak2015rfid}.

\subsection{Potential negative social impact}

Although we have tried to clear human faces and remove the majority part of human bodies in our photos using a body detection framework, this dataset still inevitably contains a small amount of human figure information (\textit{eg.} hair colour, skin color, \textit{etc.}). In a random sampling based inspection ($0.1\%$ of the dataset is sampled), there are only $0.3\%$ images contain visible human body information and these often partial information (\textit{eg.} skin colour on the forearm is the most popular one).

This dataset contains garments made of different textiles. There is a chance that if one develops a customer-facing application using this dataset might misinterpret customer figures and cause unwanted racial-related output. For instance, the system might recognize customers with coloured skin to be suede or leather. However, for textile recycling based applications, no humans are involved and this potential social problem is avoided.

\section{Data access and maintenance}

All the information about the dataset, including links to the paper, code and future announcements will be accessible at \url{https://github.com/hahasue/TextileNet/}, this information will be maintained by the authors. For images without human faces, we provide the URLs that point to the original images uploaded by either the existing dataset or google images. 
This means that for reconstructed images, the original datasets retain full control of their data - any deleted from them will be automatically removed from the URLs. 

We hosted the face-cropped images on Google Drive, users can download them by URLs or from the script. We also provide an alternative download link on One Drive. Anyone can extend TextileNet using our taxonomies and collected data with the Creative Commons Attribution (CC By) licence. The data collection code is released with an MIT license. 

In the meantime, we also release the model training code for the ResNet and Vit models on our datasets, and the pre-trained checkpoints are also included in the Github Repository.

The TexitleNet will be maintained on GitHub by the author:
\begin{itemize}
    \item GitHub for download instructions and a download script.
    \item Images with face-cropped are hosted on Google Drive and One Drive.
    \item Images that could be downloaded from other resources could be automatically downloaded in our download script.
    \item \textbf{Maintenance}: our GitHub repository will be open to the public, and we will actively try to resolve the issues raised in this repository.
\end{itemize}

\end{document}